\documentclass[fleqn,usenatbib,useAMS]{mnras}
 \usepackage{graphicx}
 \usepackage{amsmath}
 \usepackage{amssymb}
 \usepackage[T1]{fontenc}
 \usepackage{ae,aecompl}
 \usepackage{txfonts}
 \usepackage{enumerate}

 \title[Orbital evolution of meteoroid 2020~CD$_{3}$]
       {On the orbital evolution of meteoroid 2020~CD$_\mathbf{3}$, 
        a temporarily captured orbiter of the Earth--Moon system} 

 \author[C. de la Fuente Marcos and R. de la Fuente Marcos]
        {C.~de~la~Fuente~Marcos$^{1}$\thanks{E-mail: nbplanet@ucm.es}
         and
         R. de la Fuente Marcos$^{2}$ \\
         $^1$Universidad Complutense de Madrid,
             Ciudad Universitaria, E-28040 Madrid, Spain \\
         $^2$AEGORA Research Group,
             Facultad de Ciencias Matem\'aticas,
             Universidad Complutense de Madrid,
             Ciudad Universitaria, E-28040 Madrid, Spain}

 \date{Accepted 2020 March 19. 
       Received 2020 March 19; 
       in original form 2020 March 1}
 \pubyear{2020}
 
\begin{document}
  \label{firstpage}
  \pagerange{\pageref{firstpage}--\pageref{lastpage}}
  \maketitle

  \begin{abstract}
     Any near-Earth object (NEO) following an Earth-like orbit may eventually 
     be captured by Earth's gravity during low-velocity encounters. This 
     theoretical possibility was first attested during the fly-by of 1991~VG 
     in 1991--1992 with the confirmation of a brief capture episode ---for 
     about a month in 1992 February. Further evidence was obtained when 
     2006~RH$_{120}$ was temporarily captured into a geocentric orbit from 
     July 2006 to July 2007. Here, we perform a numerical assessment of the 
     orbital evolution of 2020~CD$_{3}$, a small NEO found recently that 
     could be the third instance of a meteoroid temporarily captured by 
     Earth's gravity. We confirm that 2020~CD$_{3}$ is currently following a 
     geocentric trajectory although it will escape into a heliocentric path 
     by early May 2020. Our calculations indicate that it was captured by the 
     Earth in 2016$_{-4}^{+2}$, median and 16th and 84th percentiles. This 
     episode is longer (4$_{-2}^{+4}$~yr) than that of 2006~RH$_{120}$. Prior 
     to its capture as a minimoon, 2020~CD$_{3}$ was probably a NEO of the 
     Aten type, but an Apollo type cannot be excluded; in both cases, the 
     orbit was very Earth-like, with low eccentricity and low inclination, 
     typical of an Arjuna-type meteoroid. A few clone orbits remained 
     geocentric for nearly a century, opening the door to the existence of 
     yet-to-be-detected minimoons that are relatively stable for time-scales 
     comparable to those of unbound quasi-satellites such as 
     (469219)~Kamo`oalewa 2016~HO$_{3}$. In addition, nearly 10 per cent of 
     the experiments led to brief moon-moon episodes in which the value of the 
     selenocentric energy of 2020~CD$_{3}$ became negative.
  \end{abstract}


  \section{Introduction}
     If a minor body encounters a planet at very low relative velocity, a temporary capture may occur. This theoretical possibility has been
     confirmed multiple times in the case of the outer planets (see e.g. \citealt{1981A&A....94..226C}). The Earth is not strange to this
     dynamical situation. If the orbit of a minor body is somewhat Earth-like, low-velocity encounters (as low as 0.9~km~s$^{-1}$) close or
     inside the Hill radius of the Earth, 0.0098~au, may lead to temporary capture episodes \citep{2012Icar..218..262G}. This is 
     particularly true for recurrent transient co-orbitals of the horseshoe type (see e.g. \citealt{2018MNRAS.473.2939D,
     2018MNRAS.473.3434D}). That there are circumstances under which this could occur was first confirmed during the fly-by of 1991~VG in 
     1991--1992, when this near-Earth object (NEO) experienced a temporary satellite capture by the Earth for about a month in 1992 February 
     \citep{1997CeMDA..69..119T,2018MNRAS.473.2939D}. Further evidence was obtained when 2006~RH$_{120}$ was temporarily captured into a 
     geocentric orbit from 2006 July to 2007 July \citep{2008MPEC....D...12B,2008LPICo1405.8297K,2009A&A...495..967K}. Small bodies 1991~VG 
     and 2006~RH$_{120}$ have similar sizes of about 10~m and a few metres, respectively, and they could be secondary fragments of minor 
     bodies that were originally part of the main belt and abandoned their formation region under the effect of Jupiter's gravity 
     \citep{1993Natur.363..704R,2000Icar..146..176G}. While 2006~RH$_{120}$ was identified as a temporary capture while still bound to the 
     Earth \citep{2008LPICo1405.8297K}, 1991~VG was not recognized as such until some time later \citep{1997CeMDA..69..119T}. The capture 
     episodes experienced by 1991~VG and 2006~RH$_{120}$ were also rather different. 

     Following the terminology discussed by \citet{2017Icar..285...83F}, 1991~VG was subjected to a temporarily captured fly-by because it 
     did not complete at least one revolution around the Earth when bound but 2006~RH$_{120}$ did, so it became a temporarily captured 
     orbiter. \citet{1979RSAI...22..181C} originally put forward that in order to be captured as a satellite of our planet, the geocentric 
     energy of the object must be negative disregarding any constraint on the duration of the capture event; \citet{1981A&A...102..165R} 
     argued that a true satellite has to be able to complete at least one revolution around our planet while its geocentric energy is 
     negative. Fedorets et al. \citeyearpar{2017Icar..285...83F} have predicted that 40 per cent of all captures should be temporarily 
     captured orbiters. With only two recorded instances of temporary capture, numerical predictions cannot be tested; it is therefore 
     important to identify additional examples to confirm and/or improve our current understanding of this phenomenon. A third example of a 
     meteoroid following a geocentric trajectory has been found recently in 2020~CD$_{3}$ \citep{2020MPEC....D..104R}. Here, we perform an 
     assessment of the orbital evolution of 2020~CD$_{3}$ using the available data and $N$-body simulations. As its current orbit is rather 
     chaotic and relatively uncertain, we adopt a statistical approach analysing the results of a large sample of orbits and focusing on how 
     and when 2020~CD$_{3}$ arrived to its current dynamical state. This paper is organized as follows. In Section~2, we present data and 
     methods. The orbital evolution of 2020~CD$_{3}$ is explored in Section~3. Our results are discussed in Section~4 and our conclusions 
     are summarized in Section~5.

  \section{Data and methods}
     Meteoroid 2020~CD$_{3}$ was discovered as C26FED2 by T. Pruyne and K. Wierzcho\'s on 2020 February 15 observing for the Mt. Lemmon 
     Survey in Arizona with the 1.5-m reflector + 10K CCD and it was found to be temporarily bound to the Earth \citep{2020MPEC....D..104R}. 
     The discovery MPEC states that no evidence of perturbations due to solar radiation pressure has been observed in orbit integrations, 
     and no link to a known artificial object has been found \citep{2020MPEC....D..104R}. Therefore, it has to be assumed that, as in the 
     case of 1991~VG and 2006~RH$_{120}$, 2020~CD$_{3}$ is a {\it bona fide} natural body, a very small asteroid or meteoroid with an 
     absolute magnitude of 31.7~mag (assumed $G=0.15$), which suggests a diameter in the range $\sim$1--6~m for an assumed albedo in the 
     range 0.60--0.01. Meteoroid 2020~CD$_{3}$ is probably smaller than 1991~VG and 2006~RH$_{120}$; Arecibo Observatory pointed at 
     2020~CD$_{3}$ for about 2 hours on 2020 March 6 with negative results, probably because it is too small and it was too far away to 
     detect with radar at that time (Taylor, private communication). Its most recent orbit determination is shown in Table~\ref{elements} 
     and it is based on 58 observations for a data-arc of 15~d. Its orbital elements are consistent with those of the Arjunas, a secondary 
     asteroid belt located around the path of our planet and originally proposed by \citet{1993Natur.363..704R}. The Arjunas are a loosely 
     resonant family of small NEOs, which form the near-Earth asteroid belt (see e.g. \citealt{2013MNRAS.434L...1D,2015AN....336....5D}).
%
%----------------------------------------------------------------------------------------------------------------------------------- TABLE I
%------------------------------------------------------------------------------------------------------- Orbital elements asteroids 2020 CD3
%
     \begin{table}
      \centering
      \fontsize{8}{11pt}\selectfont
      \tabcolsep 0.15truecm
      \caption{Values of the Heliocentric Keplerian orbital elements of 2020~CD$_{3}$ and their associated 1$\sigma$ uncertainties. The 
               orbit determination has been computed by S. Naidu and it is referred to epoch JD 2458906.5 (2020-Feb-27.0) TDB (Barycentric 
               Dynamical Time, J2000.0 ecliptic and equinox). Source: JPL's SBDB (solution date, 2020-Mar-03 08:13:59).
              }
      \begin{tabular}{lcc}
       \hline
        Orbital parameter                                 &   & value$\pm$1$\sigma$ uncertainty \\
       \hline
        Eccentricity, $e$                                 & = &   0.041280$\pm$0.000002         \\
        Perihelion, $q$ (au)                              & = &   0.98566379$\pm$0.00000009     \\
        Inclination, $i$ (\degr)                          & = &   0.92865$\pm$0.00007           \\
        Longitude of the ascending node, $\Omega$ (\degr) & = & 140.5181$\pm$0.0006             \\
        Argument of perihelion, $\omega$ (\degr)          & = & 338.47199$\pm$0.00008           \\
        Time of perihelion passage, $\tau$ (TDB)          & = & 2458868.5067$\pm$0.0005         \\
        Absolute magnitude, $H$ (mag)                     & = &  31.7$\pm$0.3                   \\
       \hline
      \end{tabular}
      \label{elements}
     \end{table}
%
%-------------------------------------------------------------------------------------------------------------------------------------------
%

     The orbit determination in Table~\ref{elements} is still uncertain and its associated evolution rather chaotic (see Section 3). 
     \citet{1998AJ....115.2604W} have shown that the statistical analysis of an extensive set of numerical simulations accounting for the 
     uncertainties associated with an orbit determination can produce reliable results. In order to obtain sufficiently robust conclusions, 
     we apply the Monte Carlo using the Covariance Matrix (MCCM) method detailed in section 3 of \citet{2015MNRAS.453.1288D} to generate 
     initial conditions for our $N$-body simulations that have been carried out using Aarseth's implementation of the Hermite integrator 
     \citep{2003gnbs.book.....A}; the direct $N$-body code is publicly available.\footnote{\url{http://www.ast.cam.ac.uk/~sverre/web/pages/nbody.htm}}
     The covariance matrix required to generate initial positions and velocities for 10$^{4}$ control or clone orbits has been obtained from 
     Jet Propulsion Laboratory's Solar System Dynamics Group Small-Body Database (JPL's SSDG SBDB, 
     \citealt{2015IAUGA..2256293G}).\footnote{\url{https://ssd.jpl.nasa.gov/sbdb.cgi}} 
     This is also the source of the data in Table~\ref{elements}; JPL's \textsc{horizons}\footnote{\url{https://ssd.jpl.nasa.gov/?horizons}} 
     ephemeris system \citep{GY99} has been used to gather most input data used in our calculations (e.g. data in Table~\ref{vector}). Some 
     data have been retrieved from JPL's SBDB using the tools provided by the Python package Astroquery \citep{2019AJ....157...98G,
     2019JOSS....4.1426M}. Extensive details on our calculations and physical model can be found in \citet{2012MNRAS.427..728D,
     2018MNRAS.473.2939D,2018MNRAS.473.3434D}. 

     In order to analyse the results, we produced histograms using the Matplotlib library \citep{2007CSE.....9...90H} with sets of bins of
     constant size computed using Astropy \citep{2013A&A...558A..33A,2018AJ....156..123A} by applying the Freedman and Diaconis rule 
     \citep{FD81}. Instead of using frequency-based histograms, we considered counts to form a probability density so the area under the 
     histogram will sum to one.

  \section{Orbital evolution}
     With our calculations we aimed at answering four main questions: (i) When did this capture take place? (ii) When is 2020~CD$_{3}$ 
     leaving its current geocentric path? (iii) How was the pre-capture orbit of 2020~CD$_{3}$? (iv) How diverse is the evolution in the 
     neighbourhood of the initial conditions in Table~\ref{vector}? Answering the first two questions, we can find out if 2020~CD$_{3}$ is 
     experiencing a temporarily captured fly-by or a temporarily captured orbiter episode. The third question is directly related to the 
     origin of 2020~CD$_{3}$ and the fourth question can inform us on the reliability of our results. In order to provide robust answers to 
     all these questions, we first focus on the initial conditions in the form of Cartesian state vectors and then we use the MCCM method to 
     study how the interlaced uncertainties propagate over time and how they impact our results.
%
%---------------------------------------------------------------------------------------------------------------------------------- TABLE II
%------------------------------------------------------------------------------------------------ Geometric Cartesian state vectors 2020 CD3
%
     \begin{table}
      \centering
      \fontsize{8}{11pt}\selectfont
      \tabcolsep 0.15truecm
      \caption{Cartesian state vector of 2020~CD$_{3}$: components and associated 1$\sigma$ uncertainties. Epoch as in Table~\ref{elements}. 
               Source: JPL's SBDB.
              }
      \begin{tabular}{lcc}
       \hline
        Component                         &   &    value$\pm$1$\sigma$ uncertainty                                \\
       \hline
        $X$ (au)                          & = & $-$9.249461206824366$\times10^{-1}$$\pm$2.23296462$\times10^{-7}$ \\
        $Y$ (au)                          & = &    3.829494417698888$\times10^{-1}$$\pm$8.27974473$\times10^{-8}$ \\
        $Z$ (au)                          & = &    4.825464890142035$\times10^{-3}$$\pm$2.00658446$\times10^{-7}$ \\
        $V_X$ (au\ d$^{-1}$)              & = & $-$7.033807851735909$\times10^{-3}$$\pm$2.16520750$\times10^{-8}$ \\
        $V_Y$ (au\ d$^{-1}$)              & = & $-$1.606237689633747$\times10^{-2}$$\pm$8.21345272$\times10^{-9}$ \\
        $V_Z$ (au\ d$^{-1}$)              & = &    2.735613730104734$\times10^{-4}$$\pm$2.13010688$\times10^{-8}$ \\
       \hline
      \end{tabular}
      \label{vector}
     \end{table}
%
%-------------------------------------------------------------------------------------------------------------------------------------------
%

\begin{figure*}
       \centering
        \includegraphics[width=0.49\linewidth]{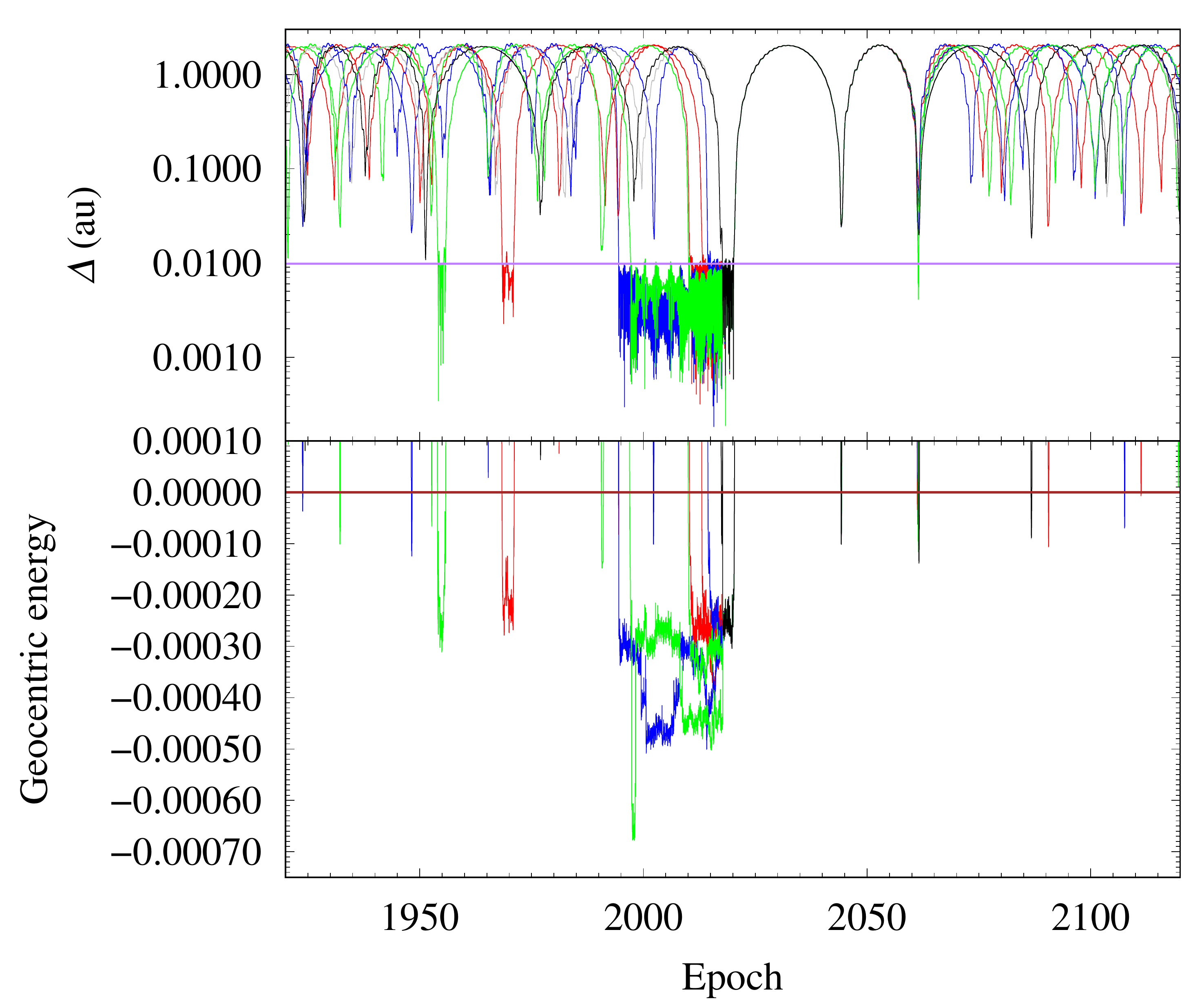}
        \includegraphics[width=0.49\linewidth]{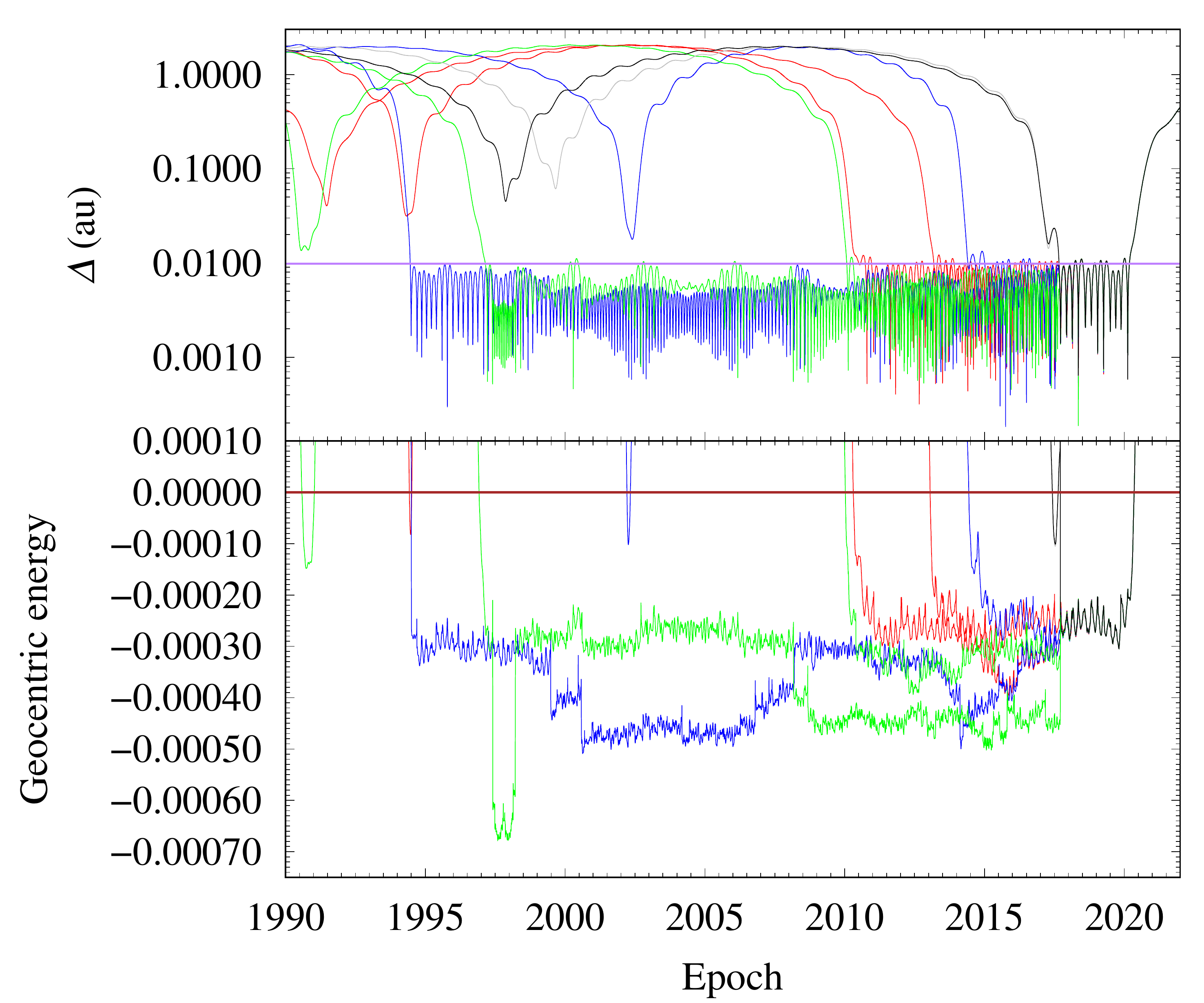}
        \caption{Evolution of the values of the geocentric distance ($\Delta$, top panel) and the geocentric energy (bottom panel) of 
                 2020~CD$_{3}$. As pointed out by \citet{1979RSAI...22..181C}, captures happen when the value of the geocentric energy 
                 becomes negative. The unit of energy is such that the unit of mass is 1~$M_{\odot}$, the unit of distance is 1~au, and the 
                 unit of time is one sidereal year divided by 2$\pi$. The right-hand side set of panels is a magnified version of those on 
                 the left-hand side. The evolution according to the nominal orbit of 2020~CD$_{3}$ in Table~\ref{elements} is shown in 
                 black, an arbitrarily close orbit appears in grey, and those of control or clone orbits with Cartesian vectors separated 
                 $\pm$1$\sigma$ (in green), $\pm$2$\sigma$ (in blue), and $\pm$3$\sigma$ (in red) from the nominal values in 
                 Table~\ref{vector}. The Hill radius of the Earth, 0.0098~au, is shown in purple (top panels).
                }
        \label{energy}
     \end{figure*}

     Figure~\ref{energy} provides an answer to the last question. It shows the short-term evolution of two important parameters ---the 
     geocentric distance, $\Delta$, and the geocentric energy--- for a reduced but relevant set of control or clone orbits. Following 
     \citet{1979RSAI...22..181C}, when the value of the geocentric energy is negative, a capture takes place. This is linked to being within 
     the Hill radius of the Earth (in purple in Fig.~\ref{energy}). The current capture episode is clearly of the temporarily captured 
     orbiter type as the value of the geocentric energy remains negative for an extended period of time, sufficient to complete multiple 
     revolutions around the Earth--Moon system as argued for by \citet{1981A&A...102..165R}. Just with the information in Fig.~\ref{energy}, 
     we cannot confirm statistically this conclusion because we are using data from Table~\ref{vector} and not taking into account how the 
     uncertainty in the value of one orbital element affects all the others. Fig.~\ref{energy} is therefore only suggestive of a 
     temporarily captured orbiter. 

     On the other hand, we observe a variety of dynamical behaviours and although those centred about the 2008 epoch are consistent with a 
     temporarily captured orbiter, temporarily captured fly-bys may have taken place in the past and they may repeat in the future (see 
     Fig.~\ref{energy}, left-hand side set of panels). This indicates that the orbital evolution of 2020~CD$_{3}$ may lead to recurrent 
     temporary captures and most episodes will be of the temporarily captured fly-by type. In addition, we observe that some initial 
     conditions tend to produce longer captures. These are associated with control or clone orbits with Cartesian vectors separated 
     $-$1$\sigma$ (in green), $+$2$\sigma$ (in blue), and $+$3$\sigma$ (in red) from the nominal values in Table~\ref{vector}. In any case, 
     the evolution is highly non-linear and a geocentric orbit of the $+$2$\sigma$ type lasts longer than those of the $-$1$\sigma$ or 
     $+$3$\sigma$ types. Gaussianly distributed initial conditions lead to rather different evolutionary outcomes due to the highly chaotic 
     dynamical context.
%
%-------------------------------------------------------------------------------------------------------------------------------------------
%
%
%-------------------------------------------------------------------------------------------------------------------------------------------
%

\begin{figure}
       \centering
        \includegraphics[width=\linewidth]{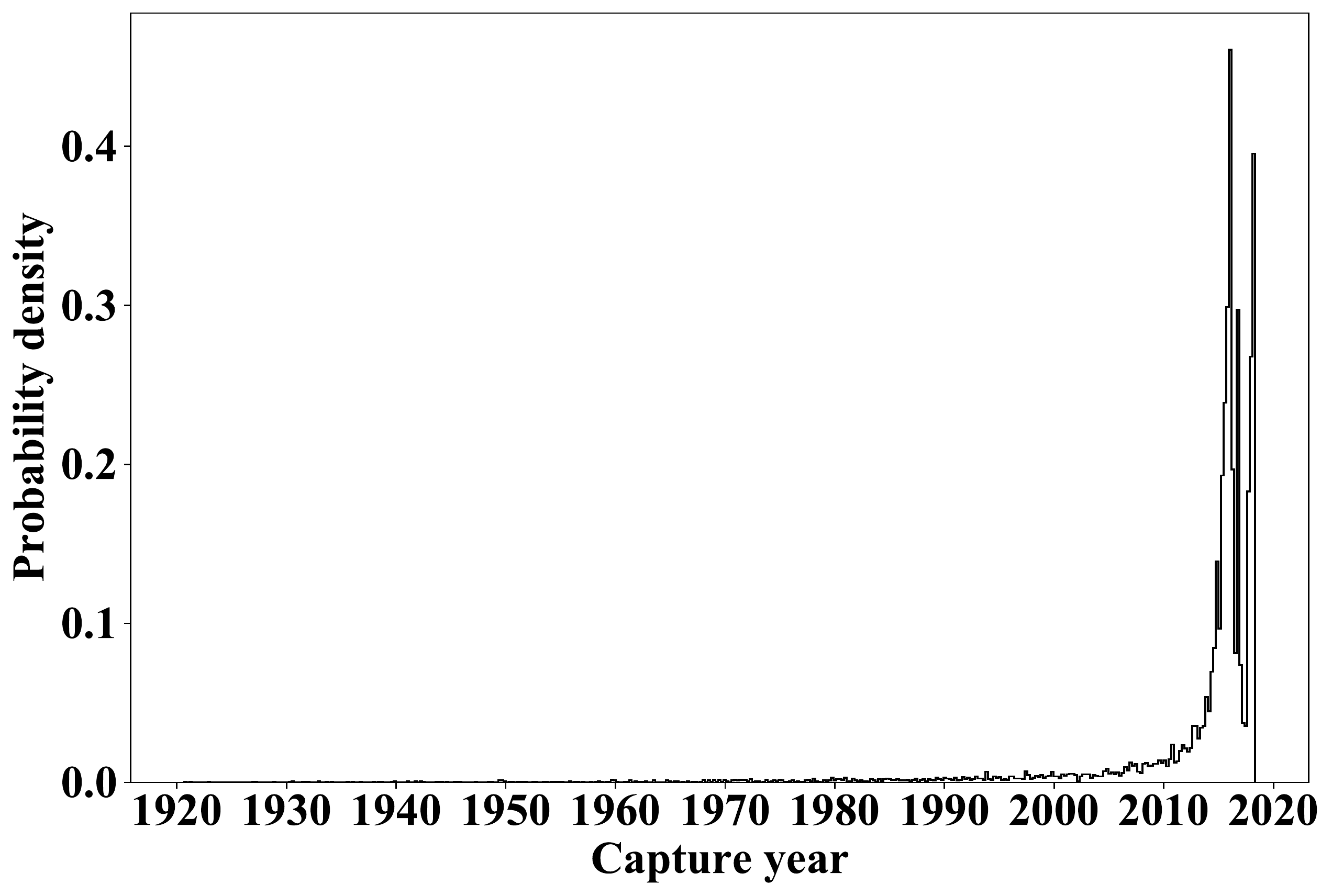}
        \caption{Distribution of the time of capture by Earth's gravity for 2020~CD$_{3}$ using initial conditions based on the orbit 
                 determination shown in Table~\ref{elements} (see the text for details). The bins in the histogram have been computed using 
                 the Freedman and Diaconis rule. 
                }
        \label{capture}
     \end{figure}

\begin{figure}
       \centering
        \includegraphics[width=\linewidth]{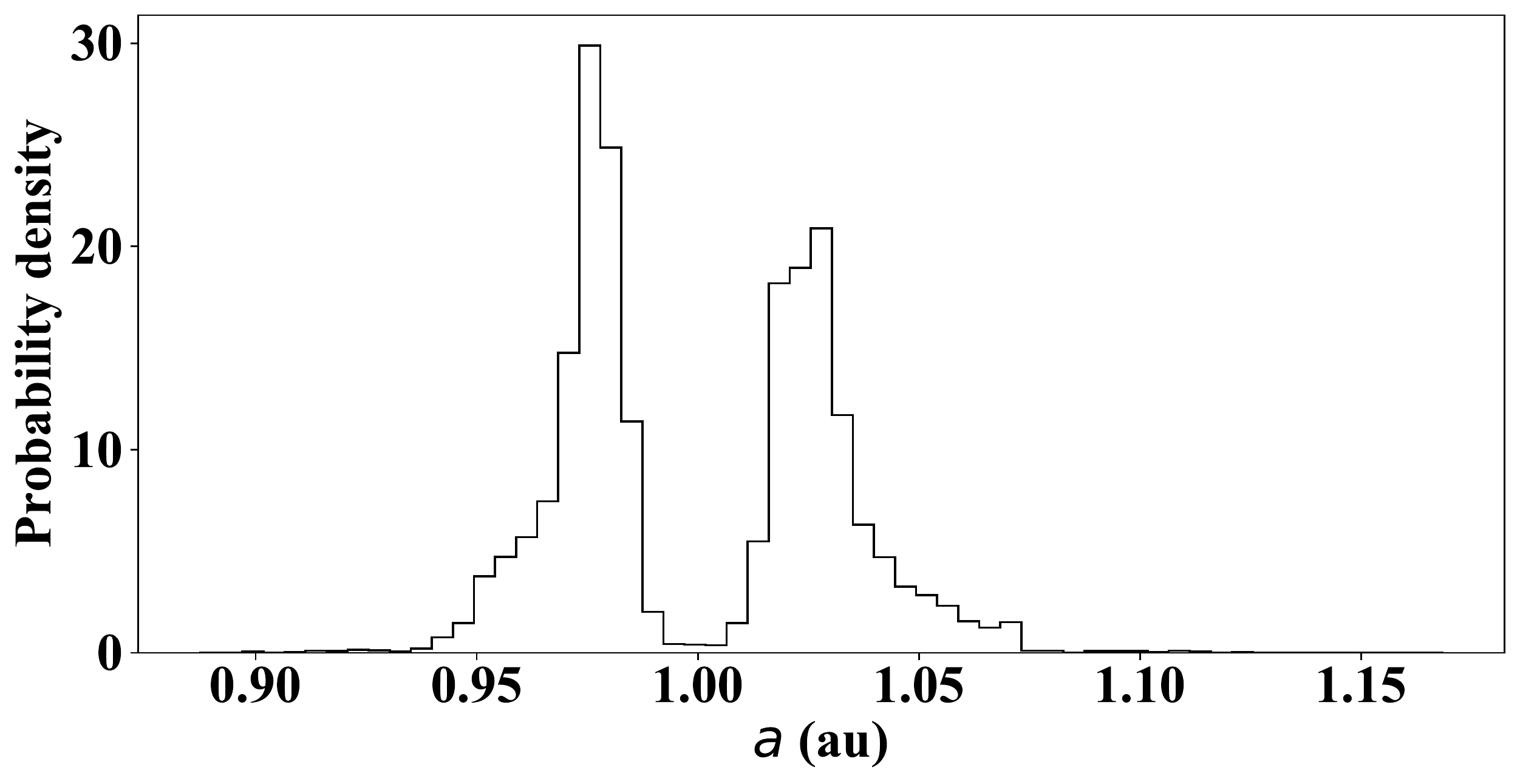}
        \includegraphics[width=\linewidth]{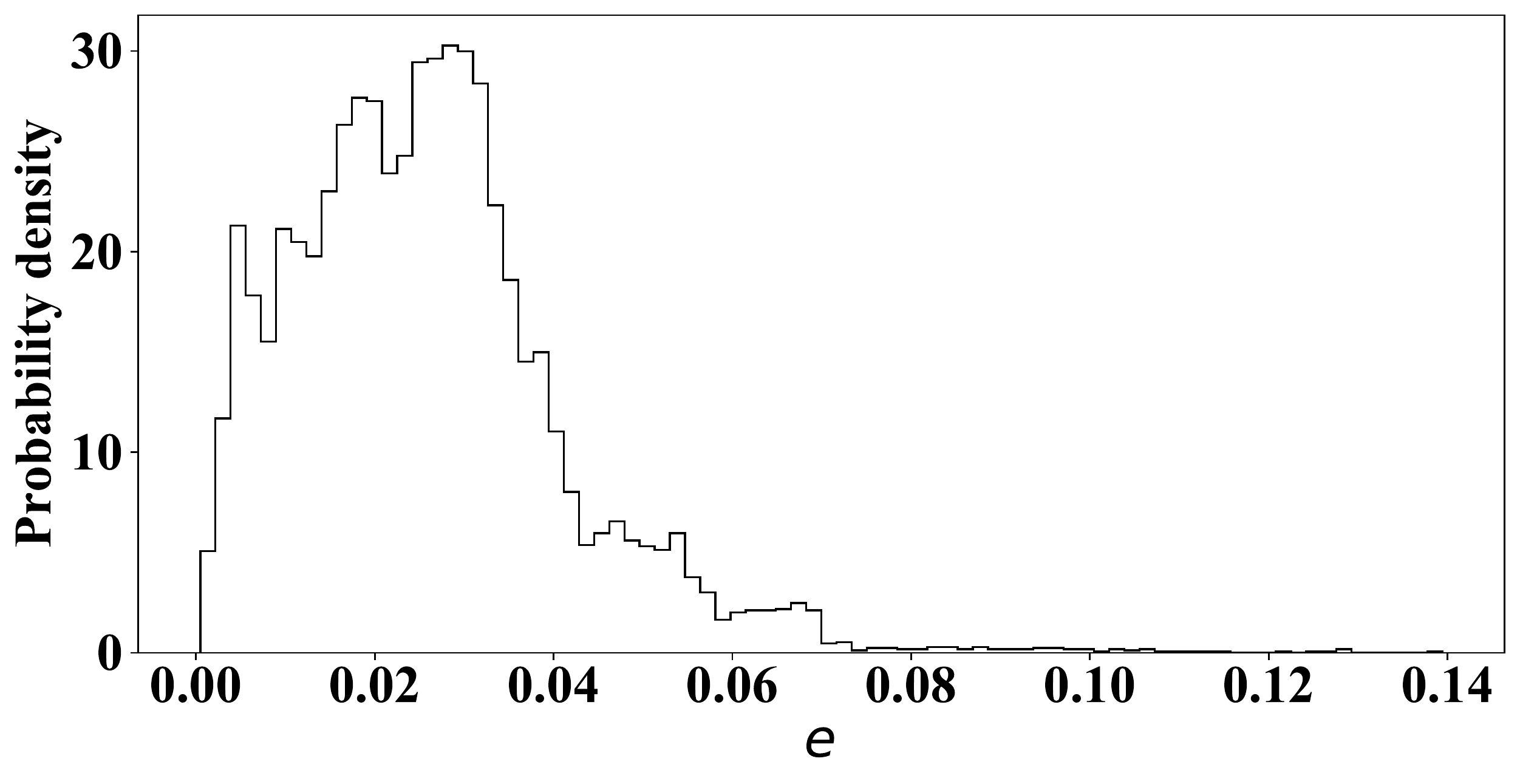}
        \includegraphics[width=\linewidth]{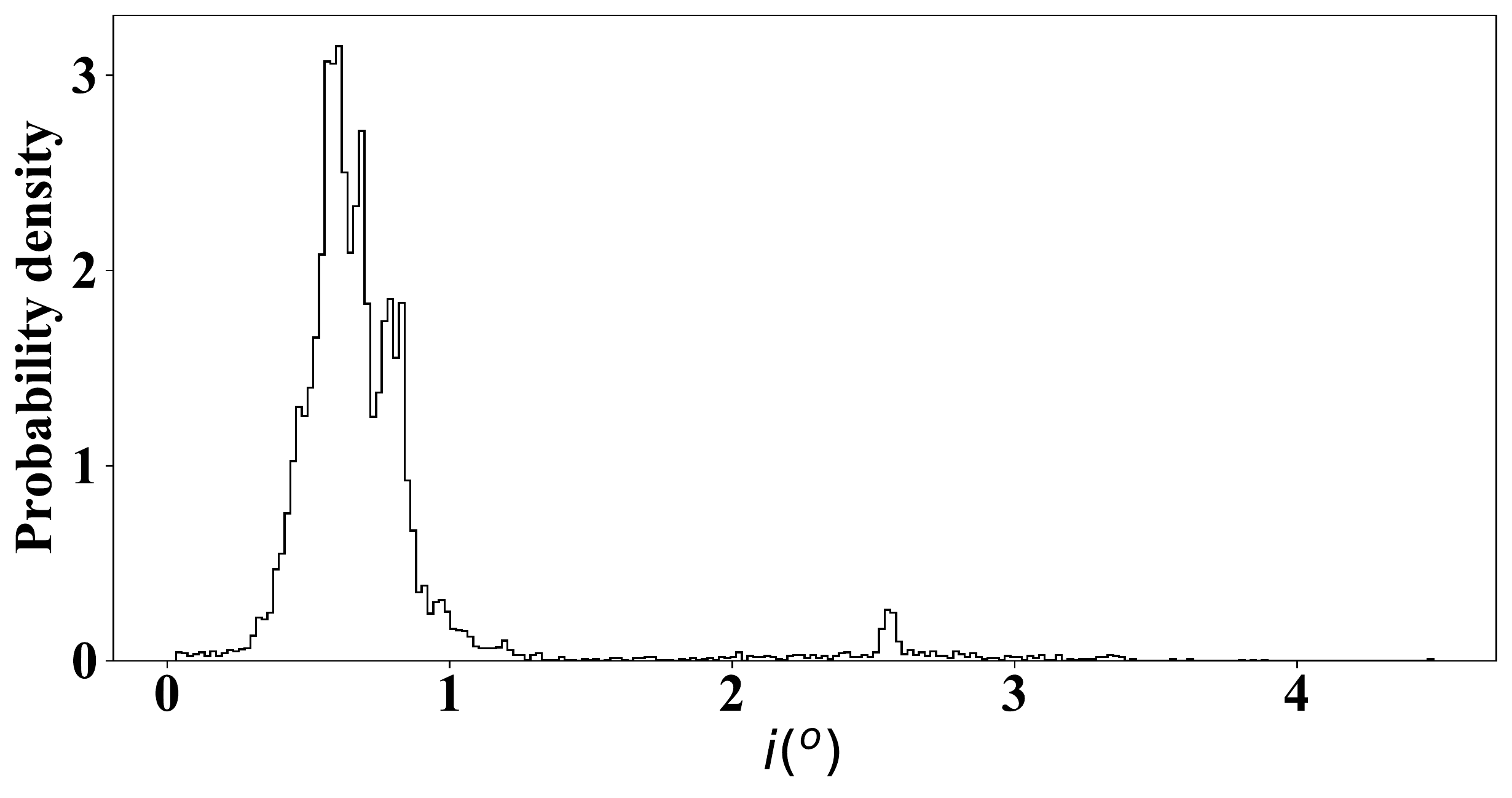}
        \caption{Distributions of semimajor axes, $a$ (top panel), eccentricities, $e$ (middle panel), and inclinations, $i$ (bottom panel), 
                 of the pre-capture orbits probably followed by 2020~CD$_{3}$ and compatible with the orbit determination in 
                 Table~\ref{elements}. The bins in the histogram have been computed using the Freedman and Diaconis rule.
                }
        \label{pre-capture}
     \end{figure}

     Figures~\ref{capture} and \ref{pre-capture} show the results of 10$^{4}$ direct $N$-body simulations for which the MCCM method has been
     used to generate initial conditions (control or clone orbits). Figure~\ref{capture} shows a trimodal distribution of the time of 
     capture. From the data in the figure, 2020~CD$_{3}$ may have been captured by the Earth on 2015.9$_{-4.4}^{+2.0}$, median and 16th and 
     84th percentiles. The outcome of the integrations backwards in time depends critically on crossing several gravitational keyholes 
     \citep{1999DPS....31.2804C} in sequence, the most important one took place late in 2017 (see Fig.~\ref{energy}, right-hand side set of 
     panels). The outcome of integrations forward in time (not shown) is far more uniform with a departure date to follow a heliocentric 
     trajectory on 2020.35109$_{-0.00011}^{+0.00010}$ (see also Fig.~\ref{energy}, right-hand side set of panels). This confirms 
     statistically that 2020~CD$_{3}$ is experiencing a temporarily captured orbiter episode, answering reliably the first two questions 
     posed above. 

     Figure~\ref{pre-capture} provides the distributions of $a$, $e$, and $i$ of the pre-capture orbits probably followed by 2020~CD$_{3}$. 
     The results are fully consistent with the expectations for an Arjuna origin, which is often linked to recurrent transient co-orbitals 
     of the horseshoe type (see e.g. \citealt{2018MNRAS.473.2939D,2018MNRAS.473.3434D}). The distribution in $a$ is bimodal 
     (Fig.~\ref{pre-capture}, top panel) reflecting the fact that it may have been an Aten (slightly more likely) or an Apollo prior to 
     capture ---switching back and forth between dynamical classes is typical of co-orbitals of the horseshoe type. The distribution in $e$ 
     (Fig.~\ref{pre-capture}, middle panel) gives a most probable value of 0.025$\pm$0.014 and that of $i$ (Fig.~\ref{pre-capture}, bottom 
     panel) has 0\fdg65$_{-0\fdg13}^{+0\fdg2}$. Meteoroid 2020~CD$_{3}$ is the only known object that may have crossed the volume of NEO 
     orbital parameter space defined by $a\in(0.95, 1.07)$~au, $e\in(0.011, 0.039)$, and $i\in(0\fdg51, 0\fdg85)$. This suggests that 
     hypothetical objects following similar orbits may have been removed from NEO space, either as a result of collisions with the Earth or 
     the Moon (see e.g. \citealt{1995Icar..118..302G,2002Natur.420..294B,2016AJ....151..135C}) or after being ejected dynamically following 
     a close fly-by. 

     The most probable values of the orbital parameters of 2020~CD$_{3}$ after escaping capture will be 
     $a=1.019297_{-0.000004}^{+0.000003}$~au, $e=0.020268_{-0.000013}^{+0.000012}$, and $i=0\fdg60900\pm0\fdg00011$, corresponding to an 
     Arjuna orbit of the Apollo type. Out of the known minor bodies, the one with the closest values to these orbital parameters is 2009~BD 
     that has $H=28.1$~mag with an area to mass ratio compatible with that of a very porous rock and for which direct detection of radiation 
     pressure effects has been reported \citep{2012NewA...17..446M}. This meteoroid may impact the Earth in the near future with impact
     solutions starting in 2071 (Micheli et al. \citeyear{2012NewA...17..446M}) and this supports our previous conclusion that objects 
     moving along similar orbits may have been removed from NEO space via collisions with the Earth or the Moon.
%
%-------------------------------------------------------------------------------------------------------------------------------------------
%
%
%-------------------------------------------------------------------------------------------------------------------------------------------
%
%
%-------------------------------------------------------------------------------------------------------------------------------------------
%
%
%-------------------------------------------------------------------------------------------------------------------------------------------
%

  \section{Discussion}
\begin{figure}
       \centering
        \includegraphics[width=\linewidth]{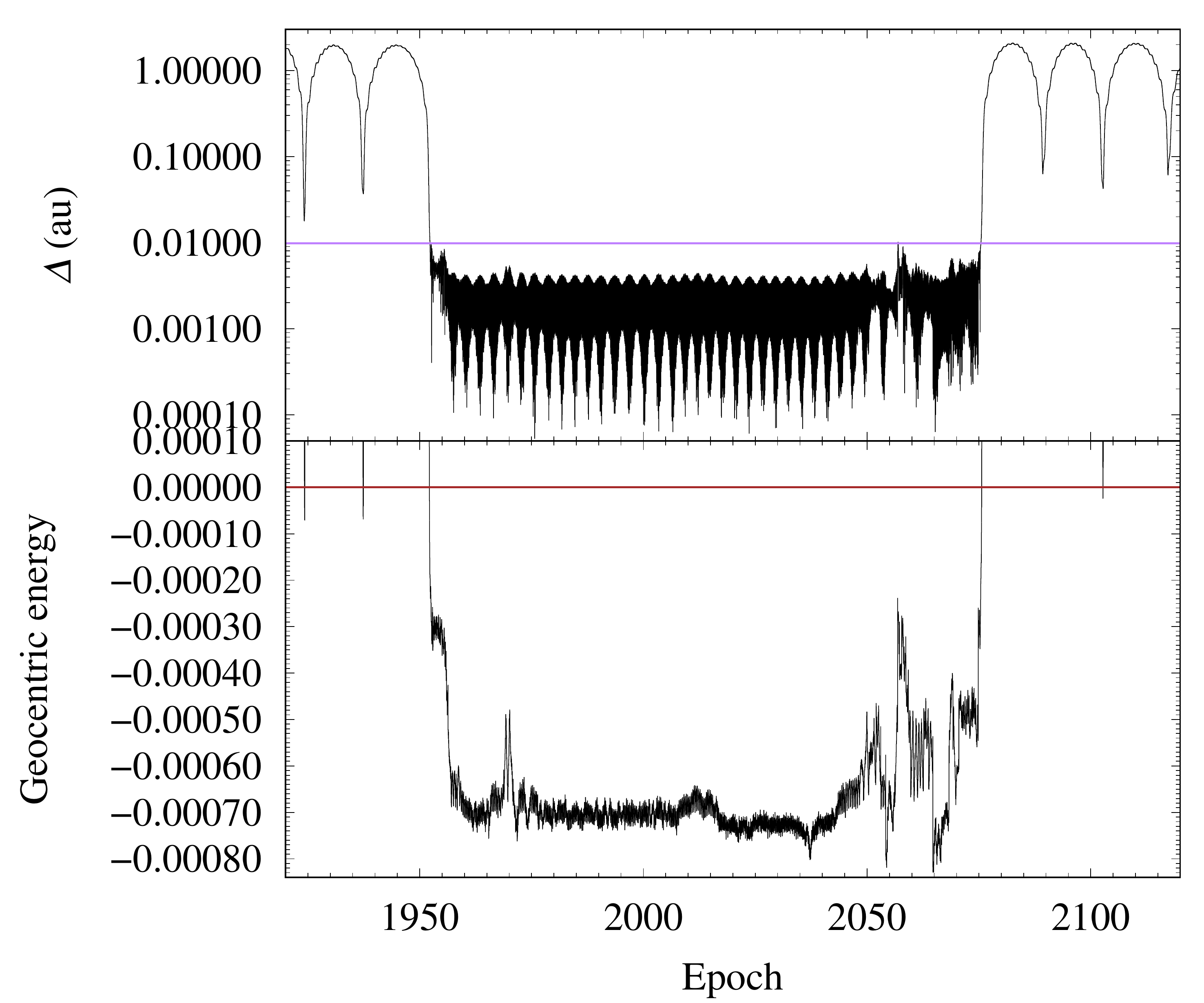}
        \caption{Similar to Fig.~\ref{energy} but for a different control or clone orbit from one of the earliest orbit determinations of 
                 2020~CD$_{3}$ available from JPL's SBDB. 
                }
        \label{energylong}
     \end{figure}

     After providing robust answers to the four questions posed at the beginning of Section~3, one may argue that other co-orbital objects 
     such as quasi-satellites and horseshoe librators \citep{1999ssd..book.....M} have a greater intrinsic interest because they remain in 
     regions relatively easy to access from the Earth during longer time-scales. For example, Earth's quasi-satellite (469219)~Kamo`oalewa 
     2016~HO$_{3}$ may remain as an unbound companion to the Earth for a few centuries \citep{2016MNRAS.462.3441D}. The fact is that some of 
     the control or clone orbits studied in our numerical survey exhibit stability for comparable time-scales. Figure~\ref{energylong} shows 
     an example in which the capture episode lasts longer than a century, opening the door to the existence of yet-to-be-detected minimoons 
     that could be relatively stable for sufficiently long periods of time.   
%
%-------------------------------------------------------------------------------------------------------------------------------------------
%
%
%-------------------------------------------------------------------------------------------------------------------------------------------
%

     On the other hand, our calculations show that the role of the gravitational attraction of the Moon on the orbital evolution of 
     2020~CD$_{3}$ is far from negligible. We have found that nearly 10 per cent of the experiments carried out led to brief (a few hours to 
     a few days) recurrent moon-moon episodes in which the value of the selenocentric energy of 2020~CD$_{3}$ became negative but failing to 
     complete one full revolution around the Moon. Following Fedorets et al. \citeyearpar{2017Icar..285...83F}, 2020~CD$_{3}$ may have had 
     temporarily captured fly-bys with the Moon. This dynamical situation was first mentioned by \citet{2019MNRAS.483L..80K} and our 
     analysis and results give some support to their conclusions.

     Although the state-of-the-art orbit model\footnote{\url{http://neo.ssa.esa.int/neo-population}} developed by the Near-Earth Object 
     Population Observation Program (NEOPOP) and described by \citet{2018Icar..312..181G} cannot make predictions regarding bodies as small
     as 2020~CD$_{3}$, the dynamical class to which it belongs, the Arjunas, cannot be fully explained within this orbit model. This is 
     probably because fragmentation processes have not been included and some Arjunas may have been locally produced via sub-catastrophic 
     impacts, tidal disruptions during very close encounters with the Earth and/or the Moon or, more likely, due to the action of the 
     Yarkovsky--O'Keefe--Radzievskii--Paddack (YORP) mechanism (see e.g. \citealt{2016MNRAS.456.2946D,2018MNRAS.473.3434D,
     2020arXiv200108786J}). Meteoroid 2020~CD$_{3}$ may have its origin in one YORP-driven fragmentation event.  
  
  \section{Conclusions}
     In this paper, we have explored the sequence of events that led to the capture of 2020~CD$_{3}$ as a minimoon of the Earth. This is
     only the second time a minor body has been discovered while still engaged in geocentric motion ---the first one was 2006~RH$_{120}$ 
     \citep{2008MPEC....D...12B,2008LPICo1405.8297K,2009A&A...495..967K}. Due to the chaotic path followed by 2020~CD$_{3}$, our 
     exploration has been carried out in statistical terms using the results of 10$^{4}$ direct $N$-body simulations. Our conclusions can be 
     summarized as follows.    
     \begin{enumerate}[(i)]
        \item All the control or clone orbits indicate that 2020~CD$_{3}$ is currently following a geocentric trajectory. This moonlet will 
              end its current capture episode on 2020 May 6--7 and it has remained as a second moon to the Earth for several 
              (4$_{-2}^{+4}$~yr) years now. Future, shorter capture events cannot be discarded.
        \item During its orbital evolution as second satellite of the Earth, 2020~CD$_{3}$ may have experienced (with a probability of 
              about 10 per cent) brief subsatellite episodes in which the value of its selenocentric energy became negative for a few hours 
              or days (i.e. a temporarily captured fly-by with the Moon). 
        \item Meteoroid 2020~CD$_{3}$ belongs to the population of NEOs that may experience recurrent transient co-orbital episodes of the 
              horseshoe or even quasi-satellite type with the Earth. It was part of this population before its capture as a minimoon and it
              will return to it after its escape from its current geocentric path. This is consistent with the analysis carried out by 
              \citet{2018MNRAS.473.2939D} for the case of 1991~VG.
     \end{enumerate}
     Spectroscopic studies carried out before it leaves the neighbourhood of the Earth--Moon system may be able to confirm if 2020~CD$_{3}$
     is indeed a natural object and provide its physical characterization.
 
     Our results suggest that, in the case of the Earth, relatively long (for a century or more) capture events are possible (see 
     Fig.~\ref{energylong}). They also show that the answer to the question posed by \citet{2019MNRAS.483L..80K} is probably in the 
     affirmative, moons can have moons. However, in the case of the Moon its submoons may have an ephemeral existence due to their chaotic 
     orbits. On the other hand, the fact is that 2020~CD$_{3}$ may have remained as a second moon of the Earth for several years, reaching 
     values of the apparent magnitude close to or below 20~mag; however, it has only been identified as such a few months before leaving its 
     geocentric path for a heliocentric one. This strongly suggests that other minimoon episodes may have been missed, neglected because 
     their associated trajectories resembled, perhaps too closely, those followed by hardware with an Earth origin. 

     Minimoons are not mere dynamical curiosities, the scientific and commercial sides of minimoons have been reviewed by 
     \citet{2018FrASS...5...13J} and its future discoverability from the Vera C. Rubin Observatory has been recently studied by 
     \citet{2020Icar..33813517F}. Minimoons are also targets of the EURONEAR survey \citep{2018A&A...609A.105V} from La Palma and their 
     windows of visibility have been discussed by \citet{2014Icar..241..280B}. 

  \section*{Acknowledgements}
     We thank the anonymous referee for her/his quick and constructive report, S.~J. Aarseth for providing the code used in this research 
     and for comments on this manuscript, O. Vaduvescu and M. Popescu for comments on observational NEO results, P.~A. Taylor for comments 
     on the radar observations carried out from Arecibo, and A.~I. G\'omez de Castro for providing access to computing facilities. This work 
     was partially supported by the Spanish `Ministerio de Econom\'{\i}a y Competitividad' (MINECO) under grant ESP2017-87813-R. In 
     preparation of this paper, we made use of the NASA Astrophysics Data System and the MPC data server. This research made use of 
     Astropy,\footnote{http://www.astropy.org} a community-developed core Python package for Astronomy \citep{2013A&A...558A..33A,
     2018AJ....156..123A}.

  \bsp
  \label{lastpage}
\end{document}